\documentclass{article}
\usepackage{array,arydshln}
\usepackage{graphicx}
\usepackage[all]{xy}
\title{Pfaffian Systems of $A$-Hypergeometric Systems II --- Holonomic Gradient Method}
\author{K.Ohara, N.Takayama}
\date{May 9, 2015}
\def\pd#1{\partial_{#1}}
\def\cc{c}
\def\pp{p}
\def\qed{ \quad Q.E.D. }
\def\comment#1{ }
\newtheorem{algorithm}{Algorithm}

\newtheorem{example}{Example}

\newtheorem{theorem}{Theorem}
\begin{document}
\maketitle


\section{Introduction}

Let $A$ be $d \times n$ matrix with entries in 
${\bf N}_0 = \{ 0, 1, 2, \ldots \}$.
The $i$-th column of the matrix $A$ is denoted by $a_i$.
For a given $\beta \in {\bf N}_0 A = \{ \sum_{i=1}^n v_i a_i \,|\, v_i \in {\bf N}_0 \}$,
we consider the set of points
$$ {\cal F}(\beta) =  \{  u \in {\bf N}_0^d \,|\, A u = \beta \}. $$
We are interested in the probability distribution
$\frac{1}{u!} p^u/Z(\beta;p)$
on ${\cal F}$ 
where $u \in {\cal F}$, $p \in {\bf R}^n$ and
$Z$ is the normalizing constant defined by
$$ Z(\beta;p) = \sum_{u \in {\cal F}} \frac{1}{u!} p^u, \quad
 u! = \prod_{k=1}^n u_k!. $$
When $a_i$ lies on a hyperplane $h(y)=1$, 
the probability distribution can be regarded as
a conditional probability distribution 
induced from the multinomial distribution
$\frac{|u|!}{u!} p^u$, $ \sum_{k=1}^n p_i = 1$.
We will call this probability distribution $A$-distribution
following the celebrated work by P.Diaconis and B.Sturmfels
on constructions of  Markov bases
of ${\cal F}$ in terms of  Gr\"obner bases of the affine toric ideal
defined by the matrix $A$ \cite{DS}, \cite{dojo}.
The class of $A$-distributions includes  hypergeometric
distributions on contingency tables with a fixed marginal sums.

We will discuss on an exact evaluation of the normalizing
constant $Z$ and its derivatives in this paper.
The normalizing constant $Z$ satisfies the $A$-hypergeometric
system, which is substantially introduced by Gel'fand, Kapranov, Zelevinsky
in the late 1980's  (see, e.g., \cite{SST}),
and contiguity relations (recurrence relations) such as
$ \frac{\partial Z}{\partial p_i}(\beta;p) = Z(\beta-a_i,p)$.
By the system and contiguity relations, 
an discrete analogy of the holonomic gradient method (HGM)
introduced in \cite{N3OST2}  will give an efficient and exact method 
to evaluate the normalizing constant and its derivatives.
For the matrix $A$ standing for the $2 \times n$ contingency tables,
the maximal likelihood estimate (MLE) by the HGM is discussed 
in the recent thesis by M.Ogawa \cite{ogawa}.
The MLE needs evaluations of the normalizing constant and its derivatives,
which is done by contiguity relations for $2 \times n$ 
contingency tables given in \cite{goto}.

In this paper, we will discuss on the discrete HGM for a general matrix 
$A$.
The key step for it is a construction of the Pfaffian system 
for the $A$-hypergeometric system.
We utilize the result of \cite{HNT}, 
Macaulay type matrices \cite{Macaulay}, 
and the Hilbert driven Buchberger algorithm \cite{traverso}
for an efficient construction.
Timing data are given to compare several methods.

\section{Macaulay Type Matrices} \label{sec:macaulay}

The discussion of this section is well-known, 
but we do not find 
a relevant literature for our application.
Then, we give a brief explanation on Macaulay type matrices
(which has an origin in \cite[p.8]{Macaulay}).
Let $P=K[x_1, \ldots, x_n]$ be the ring of polynomials of $n$ variables
where $K$ is a coefficient field.
We suppose that a zero-dimensional ideal $I$ of $P$ 
is generated by  polynomials
$f_1, \ldots, f_m$.
Let $g$ be an element of $I$.
Then, $g$ can be written as 
\begin{eqnarray} \label{eq:linrel}
g = \sum_{j} \left( \sum_i c_{ij} t_{ij}\right) f_j = \sum_{ij} c_{ij} t_{ij} f_j
\end{eqnarray}
where $c_{ij} \in K$ and $t_{ij}$ is a monomial.
Let $S$ be a basis, consisting monomials, 
of $P/I$ as a vector space over $K$.
$S$ is a finite set.
We assume that a base $S$ of $P/I$ is given.
In fact, in our application to $A$-hypergeometric systems,
the base $S$ is determined by an algorithmic method in \cite{HNT}
without computing a Gr\"obner basis for a given $A$-hypergeometric ideal.
We define ${\tilde M}$ the union 
with respect to $i$'s and $j$'s 
of the set of the monomials which appear in $t_{ij} f_j$.
Set $M = {\tilde M} \setminus S$.
Let us construct a matrix $F$ of which columns are indexed by $M \cup S$
and whose each row stands for the coefficients of $t_{ij} f_j$.
We denote the row consisting of the coefficients of $t_{ij} f_j$ by $C(t_{ij} f_j)$.
For $g$ in (\ref{eq:linrel}),
$C(g)$ is the row vector of the coefficients of the polynomial $g$,
which is simplified,
indexed by $M \cup S$.
It follows from the relation (\ref{eq:linrel}) that
the vector 
$C(g)$ can be written as a linear combination of the rows of the matrix $F$.

Let us present an example to illustrate our definitions.
Put $f_1 = x_1^2+x_2^2-4$, $f_2=x_1 x_2 -1$.
Let $<$ the lexicograhic order such that $x_1 > x_2$.
Then, the set of the standard monomials $S$ is 
$\{  x_2^3, x_2^2,x_2,1\}$.
Put $g = x_2 f_1 - x_1 f_2 = x_1 + x_2^3-4 x_2$.
Then, $M=\{ x_1^2 x_2, x_1 \}$. 
we  have
$$ 
\bordermatrix{
M \cup S & x_1^2 x_2 & x_1 & &  x_2^3 & x_2^2 & x_2 & 1 \cr
C(x_2f_1)  &  1          & 0   & &  1      &  0     &-4   & 0 \cr
C(x_1f_2)  &  1          & -1 & &  0      &  0     & 0    & 0 \cr
C(g)        &  0          & 1   & &  1      &  0     &-4   & 0 \cr       
}
$$
and
$$ F =
\bordermatrix{
M \cup S & x_1^2 x_2 & x_1 & &  x_2^3 & x_2^2 & x_2 & 1 \cr
C(x_2f_1)  &  1          & 0   & &  1      &  0     &-4   & 0 \cr
C(x_1f_2)  &  1          & -1 & &  0      &  0     & 0    & 0 \cr
}.
$$

We suppose that the support of $g$, which is an element of $I$ expressed as (\ref{eq:linrel}), 
is $S$ and a monomial $t$ in $M$.
We suppose that the coefficient of $t$ in $g$ is normalized to be $1$.
The monomial $t$ can be uniquely written as a linear combination
of the elements of $S$ over $K$ modulo $I$ as
$ t = \sum_{s \in S} c_s s \  {\rm mod}\, I$, $c_s \in K$.
Then we have $g = t-\sum_{s \in S} c_s s \in I$. 
Since $C(g)$ is in the row span of $F$, we can obtain $C(g)$ 
by constructing reduced row echelon form of the matrix 
$F$ of $C(t_{ij} f_j)$'s 
with the index order $M > S$.
In fact, let $e$ be the number of echelons (the number of
the leading terms of the reduced Gr\"obner basis of the system of
linear equations defined by the matrix $F$).
Then, the rank of $F$ is equal to $e$.
If $C(g)$ is not contained in the reduced row echelon form,
then the rank of 
$\left( \begin{array}{c} F \\ C(g) \\ \end{array} \right)$
is $e+1$.
It contradicts to the assumption that $C(g)$ is contained in the row span
of $F$.

For polynomials $f_1, \ldots, f_m$,
a positive number $T$, 
and a basis $S$ of $P/I$ consisting of monomials,
we construct 
a matrix by row vectors of the form $C(t f_i)$, $i=1, \ldots, m$
where $t$ runs over the set of the monomials of which degrees 
are less than or equal to $T$.
We will call the matrix the {\it Macaulay type matrix} 
of degree $T$ for $f_1, \ldots, f_m$ and $S$.
We denote the matrix by $F_T$.

\section{Macaulay Type Matrices to Derive Pfaffian Systems}

The method finding $C(g)$ by computing the reduced row echelon form from the Macaulay type matrix
has no advantage to Gr\"obner basis methods 
without new ideas as in the $F_4$ algorithm and its related algorithms,
however, as we will see in this paper, this method has an advantage
for the numerical analysis of $A$-hypergeometric systems
in the ring of differential operators or in the ring of
difference-differential operators without them.
Consider the ring of differential operators
with rational function coefficients 
$R'=K'\langle \pd{1}, \ldots, \pd{n} \rangle$
where $K'={\bf C}(x_1, \ldots, x_n)$, which is abbreviated as ${\bf C}(x)$.
Let $I$ be a zero-dimensinal left ideal in $R'$,
$S$ a basis of $R/I$ over $K'$.
Let $s$ be an element of $S$.
Then, $\pd{i} s $ is uniquely written as the linear combination of $S$
over $K'$ modulo $I$
as $\pd{i} s = \sum_{ t \in S} c_t t \ {\rm mod}\, I$, $c_t \in K$.
Let us denote by $g$ the expression $\pd{i} s - \sum_{t \in S} c_t t$.
We call $g$ the {\it Pfaffian operator} for $\pd{i}$, $s$ and $S$.
We want to find all Pfaffian operators for all pairs of $\pd{i}$,
$i=1, \ldots, n$  and $s\in S$.
We utilize Macaulay type matrices explained in the section \ref{sec:macaulay},
which can be defined analogously in $R'$.
Let $F=F_T$ be the Macaulay type matrix constructed from $t_{ij}$'s whose total 
degrees are less than or equal to $T$ and $f_j$'s which are generators of  $I$.
When the total degree $T$ is sufficiently large, then 
the ideal element $g$, 
of which support is in $S \cup \{ \pd{i} s \}$,
can be obtained by computing the reduced row echelon form of $F$
with an index order such that $M > S$ 
and $\pd{i} s$ is the smallest index in $M$.
We note that the  linear algebra elimination over $K'$
to obtain the reduced row echelon form does not perform
differentiations of  elements of $K'={\bf C}(x)$ and then
the numerical coefficients of $g$ when the variable $x$ is specialized
to a generic constant vector $X$ can be obtained by specializing
the matrix $F$ to $x=X$ and by the linear algebra elimination.
This will be the key idea to apply Macaulay type matrices
to evaluate $A$-hypergeometric polynomials,
in which we avoid a Gr\"obner basis computation.

Let $A$ be a $d \times n$ matrix.
The $i$-th column of $A$ is denoted by $a_i$
and the $(i,j)$ element of $A$ is denoted by $a_{ij}$.
We assume that $a_i \in {\bf N}_0^d$ and
the rank of $A$ is $d$.
Let us consider Macaulay type matrices when $I$ is the $A$-hypergeometric ideal
$H_A(\cc)$.
We regard the parameters $\cc_i$'s as  indeterminates, and
we put 
$$K={\bf C}(\cc_1,\ldots, \cc_d,x_1, \ldots, x_n)
\ \mbox{ and }  \ 
R = K\langle \pd{1}, \ldots, \pd{n} \rangle.
$$
Generators of $H_A(\cc)$ consist of
the Euler operators 
$E_j-\cc_j = \sum_{k=1}^n a_{jk} x_k \pd{k}-c_j$, $j = 1, \ldots, d$ and
the toric ideal 
$I_A = \langle \pd{}^u - \pd{}^v \,|\, Au=Av, u,v  \in {\bf N}_0^n \rangle$.
It is important to distiguish the left ideal $RH_A(c)$ and
the left ideal $R'H_A(\beta)$, $\beta \in {\bf C}^d$.
The former ideal lies in $R$ where $c_i$'s are indeterminates
and the latter ideal lies in $R'$.
We consider a Macaulay type matrix  for $H_A(\cc)$.
We will construct a smaller matrix $F'$
than the matrix $F$ 
to obtain Pfaffian operators 
in the following way.

\comment{prime ${}'$  is used for smaller set.} 
For a given term order $<$,
the reduced Gr\"obner basis of $I_A$ with respect to the order $<$ 
is denoted by $G$.
A basis of  $R/RH_A(c)$ can be found by the algorithm in \cite{HNT}
from $G$.
We denote it by $S$.
Take a monomial $s$ from the set $S$.
\begin{algorithm} \rm \quad \\   \label{alg:MacaulayTypeMatrix}
Input: a matrix $A$ which defines an $A$-hypergeometric system and a sufficiently large integer $T$. \\
Output: A matrix $F'_T$ with entries in $K$, which will be called
the {\it Macaulay type matrix of degree $T$ of the $A$-hypergeometric system}.
\begin{enumerate}
\item Compute a Gr\"obner basis $G$ of $I_A$.
\item Find a basis $S$ by the algorithm in \cite{HNT}.
(A probabilistic version of this method in \cite{HNT2} can be used
and is more efficient.) 
\item Let $N_T$ be the set of the monomials 
in $\pd{i}$'s whose total order
is less than or equal to $T$.
\item Multiply an element $t$ of $N_T$ to an Euler operators $E_j-\cc_j$ in 
the ring of differential operators $R$.
Reduce the element $t (E_j-\cc_j)$ by $G$ and obtain the remainder.
By the remainders from all the pairs from $N_T$ and the Euler operators,
we construct the matrix $F'$ as in the section \ref{sec:macaulay}
and set it $F'_T$.
\end{enumerate}
\end{algorithm}

\begin{theorem}  \label{th:echelon}
\begin{enumerate}
\item When $T$  is sufficiently large, the reduced row echelon form of 
$F'=F'_T$ contains
the Pfaffian operator $\pd{i} s - \sum_{t \in S} c_t t$ 
if $\pd{i}s$ is irreducible by $G$.
\item 
Let $f$ be a solution of $H_A(\cc)$.
The numerical value $(\pd{i}s) \bullet f$ at a generic point $x=X \in {\bf Q}^n$,
$\cc={\beta} \in {\bf Q}^d$,
can be obtained from the numerical values of $s \bullet f$, $s \in S$
at a point $x=X, \cc={\beta}$
by computing the reduced row echelon form of
the numerical matrix $F'|_{x=X,\cc={\beta}}$.
\end{enumerate}
\end{theorem}
We note that
when $\pd{i}s$ is reducible by $G$, it is equal to 
an irreducible monomial $\pd{}^v$ modulo $G$.
When $T$ is sufficiently large $\pd{}^v -\sum_{t \in S}c_t t$ will be in the reduced row echelon form of $F'$.
Then, we can obtain the Pfaffian operator $\pd{i} s - \sum_{t \in S}c_t t$
from the reduced row echelon form.

\medbreak

{\it Proof}\/.
1. The set $S$ is a basis of $R/RH_A(\cc)$ over 
$K={\bf C}(\cc_1,\ldots, \cc_d,x_1, \ldots, x_n)$.
We multiply $N_T$ from the left to the generatros $E_j-\cc_j$ and $\pd{}^{u}-\pd{}^v$
of $H_A(\cc)$.
${\tilde M}$ is the set of the momonials in the obtained elements of $R$
from $N_T \cdot (E_j - \cc_j)$
and $N_T \cdot G$.
Put $M = {\tilde M} \setminus S$ as in the section \ref{sec:macaulay}.
The set $G$ is a Gr\"obner basis of the toric ideal $I_A$.
We devide $M$ into two disjoint sets $M_r$ and its complement $M_r^c$
where the set $M_r$ is the set of reducible monomials by $G$.
We define the set $M_i$  
the union of $M_r^c$ and the set of irreducible monomials
obtained by reducing elements of $M_r$ by $G$,
which is denoted by ${\rm Red}(M_r)$,  minus the set $S$.
Thus, for $\pd{}^u \in M_r$, there exists $\pd{}^v \in M_i \cup S$ such that
\begin{equation}  \label{eq:toric1}
\pd{}^u -\pd{}^v \in I_A.
\end{equation}
The subset of $M_i$,
of which element appears in the column index of $F'$, is denoted by $M'$.
The set $M \cup ({\rm Red}(M_r) \setminus S)$
consists of 
the disjoint sets $M_r$, $M_i \setminus M'$, $M'$.
Construct the Macaulay type matrix $F$ from these binomials (\ref{eq:toric1}) and
the obtained elements by the multiplication of $N_T$
to the generators of $H_A(c)$.
We sort the columns of the matrix by the order $M_r > (M_i \setminus M') >M' > S $.
When $T$ is sufficiently large, the reduced row echelon form of the Macaulay type matrix $F$
contains the Pfaffian operator $\pd{i} s - \sum_{t \in S} c_t t$
as we have seen in the previous section.
The reduced row echelon form of the smaller matrix $F'$ is contained in the reduced row echelon form of $F$. \comment{in the sense of the reduced Gr\"obner basis of linear equations.}
Since $\pd{i} s \in M_i \cap M'$ for sufficiently large $T$, 
the Pfaffian operator is contained in the reduced row echelon form of $F'$.

2. The steps to find the reduced row echelon form do not have steps of multiplication in the ring of differential
operators $R$. 
Then, a specialization of $x$ and $\cc$ to numbers 
and constructing the reduced row echelon form commute.
\qed

\begin{example}\rm  \label{ex:matrixE22}
Consider the $A$-hypergeometric ideal generated by
\begin{eqnarray*}
&&
  x_1 \pd{1}+x_2 \pd{2} + x_3 \pd{3} + x_4 \pd{4} -\cc_1,
  x_2 \pd{2} + x_4 \pd{4} -\cc_2,
  x_3 \pd{3} + x_4 \pd{4} -\cc_3, \\
&& \underline{\pd{2}\pd{3}}-\pd{1}\pd{4}.
\end{eqnarray*}
For the reverse lexicographic order such that
$\pd{1}>\pd{2}>\pd{3}>\pd{4}$,
the basis 
by the algorithm \cite{HNT}
is $S=\{1, \pd{4}\}$.
Put the degree $T=1$.
We multiply the  monomials in
$N_T=\{1, \pd{1}, \pd{2}, \pd{3}, \pd{4}\}$ to the generators.
The table below is the result of the multiplication
where the index
$i_1 i_2 \cdots i_m$ in the top row 
stands for the monomial $\prod_{k=1}^m \pd{i_k}$
and the index $0$ denotes the monomial $1$.
$$
{\tiny
\begin{array}{ccccc|c|ccccccc|cc}
M'  &  &  &  &  &M_r  &M'  &  &  &  &   &   &   &S   &\\
11  &12  &13  &14  &22  &23  &24  &33  &34  &44  &1   &2   &3   &4   &0\\\hline
    &    &    &    &    &    &    &    &    &    &x_1 &x_2 &x_3 &x_4 &-\cc_1\\
x_1 &x_2 &x_3 &x_4 &    &    &    &    &    &    &1-\cc_1&    &    &    &  \\    
    &x_1 &    &    &x_2 &x_3 &x_4 &    &    &    &    &1-\cc_1&    &    &  \\
    &    &x_1 &    &    &x_2 &    &x_3 &x_4 &    &    &    &1-\cc_1&    & \\
    &    &    &x_1 &    &    &x_2 &    &x_3 &x_4 &    &    &    &1-\cc_1& \\\hdashline
    &    &    &    &    &    &    &    &    &    &    &x_2 &    &x_4 &-\cc_2\\
    &x_2 &    &x_4 &    &    &    &    &    &    &-\cc_2&    &    &    &  \\    
    &    &    &    &x_2 &    &x_4 &    &    &    &    &1-\cc_2&    &    &  \\
    &    &    &    &    &x_2 &    &    &x_4 &    &    &    & -\cc_2&    & \\
    &    &    &    &    &    &x_2 &    &    &x_4 &    &    &    &1-\cc_2& \\\hdashline
    &    &    &    &    &    &    &    &    &    &    &    &x_3 &x_4 &-\cc_3\\
    &    &x_3 &x_4 &    &    &    &    &    &    &-\cc_3&    &    &    &  \\    
    &    &    &    &    &x_3 &x_4 &    &    &    &    &-\cc_3&    &    &  \\
    &    &    &    &    &    &    &x_3 &x_4 &    &    &    &1-\cc_3&    & \\
    &    &    &    &    &    &    &    &x_3 &x_4 &    &    &    &1-\cc_3& \\\hdashline
    &    &    &-1  &    &1   &    &    &    &    &    &    &    &    & 
\end{array}
}
$$
and
$$
{\tiny
\begin{array}{c|c|ccc|ccc}
M_i\setminus M'&M_r&M_i\setminus M'&&&M_r&& \\
114&123&124&134&144&223&233&234 \\\hline
-1 &1  &   &   &   &   &   &    \\
   &   &-1 &   &   &1  &   &    \\
   &   &   &-1 &   &   &1  &    \\
   &   &   &   &-1 &   &   &1   
\end{array}
}
$$
We put 
\begin{eqnarray*}
M_t &=& \{23,114,123,124,134,144,223,233,234\} \\
M' &=& \{1,2,3,11,12,13,14,22,24,33,34,44\}.
\end{eqnarray*}
Note that 
$M_t = M_r \cup (M_i \setminus M')$
in the proof.
The join of the two tables is the matrix $F$ and
$M=M_t \cup M'$.
Let us use the order of indices such that $M_t > M' > S$.
Apply the Gaussian elimination with this order to the first table.
We eliminate elements of the column standing for the index $23$
by the last row and then remove the last row. 
Then, we obtain the following matrix which agrees with the matrix  $F'_T$,
($T=1$)
in the algorithm.
$$
{\tiny
\begin{array}{ccccc|c|ccccccc|cc}
11  &12  &13  &14  &22  &23  &24  &33  &34  &44  &1   &2   &3   &4   &0\\\hline
    &    &    &    &    &    &    &    &    &    &x_1 &x_2 &x_3 &x_4 &-\cc_1\\
x_1 &x_2 &x_3 &x_4 &    &    &    &    &    &    &1-\cc_1&    &    &    &  \\    
    &x_1 &    &-x_3&x_2 &    &x_4 &    &    &    &    &1-\cc_1&    &    &  \\
    &    &x_1 &-x_2&    &    &    &x_3 &x_4 &    &    &    &1-\cc_1&    & \\
    &    &    &x_1 &    &    &x_2 &    &x_3 &x_4 &    &    &    &1-\cc_1& \\\hdashline
    &    &    &    &    &    &    &    &    &    &    &x_2 &    &x_4 &-\cc_2\\
    &x_2 &    &x_4 &    &    &    &    &    &    &-\cc_2&    &    &    &  \\    
    &    &    &    &x_2 &    &x_4 &    &    &    &    &1-\cc_2&    &    &  \\
    &    &    &-x_2&    &    &    &    &x_4 &    &    &    & -\cc_2&    & \\
    &    &    &    &    &    &x_2 &    &    &x_4 &    &    &    &1-\cc_2& \\\hdashline
    &    &    &    &    &    &    &    &    &    &    &    &x_3 &x_4 &-\cc_3\\
    &    &x_3 &x_4 &    &    &    &    &    &    &-\cc_3&    &    &    &  \\    
    &    &    &-x_3&    &    &x_4 &    &    &    &    &-\cc_3&    &    &  \\
    &    &    &    &    &    &    &x_3 &x_4 &    &    &    &1-\cc_3&    & \\
    &    &    &    &    &    &    &    &x_3 &x_4 &    &    &    &1-\cc_3& \\
\end{array}
}
$$
We can see, by a calculation, that this matrix can be transformed into
the reduced row echelon form whose rank is 12.
The reduced row echelon form contains the Pfaffian operators. 
\comment{14,24,34,44,1,2,3,4 can be expressed in terms of 0 and 4.}
\end{example}

\comment{
Memo:
The remainder $f'$ is written as
$f'=N_T (E_j-\cc_j) - \sum c_{Tji} t_i $, $t_j \in G$.
Original $F$ (including toric ideal) is denoted by ${\hat F}$.
Procedure to find an element of Pfaffian $N_G(\pd{i}s)-\mbox{sum of std}$.
}


\section{Evaluation of $A$-Hypergeometric Polynomials by Macaulay type matrices}

Let $R'=K'\langle \pd{1}, \ldots, \pd{n}\rangle$,
$K'={\bf Q}(x_1, \ldots, x_n)$
be the ring of differential operators with rational function
coefficients.
Let $I$ be a zero dimensional left ideal of $R'$ and $S=\{s_1, \ldots, s_r\}$ 
a basis of $R/I$
as a vector space over $K'$.
We assume that the set $S$ consists of  monomials of $\pd{}$.
If no confusion arises,  the column vector
$(s_1, \ldots, s_r)^T$ is also denoted by $S$.
There exists a matrix $P_i$ with entries in $K'$ such that 
$\pd{i} S - P_i S = 0 $ modulo $I$.
The system of differential equations 
$\pd{i} Y  - P_i Y =0$,
$i=1, \ldots, r$ 
where $Y$ is a column vector of unknown functions
is called a {\it Pfaffian system}.
A basis $S$ and the  matrix $P_i$ can be obtained by computing
a Gr\"obner basis of $I$ in the ring of differential operators $R'$
(see, e.g., \cite[Chapter 6]{dojo}). 
Let $H_A(\beta)$ be the $A$-hypergeometric ideal.
Let $Z(\beta;x)$ be an $A$-hypergeometric series
for the matrix $A$ and a generic parameter vector $\beta$.
It is well known that a contiguity relation
$$ \pd{i} \bullet Z(\beta;x) = Z(\beta-a_i;x) $$
holds under a suitable normalization of the hypergeometric series,
see, e.g., \cite[p.xy]{SST}.
In particular, the relation holds when $Z$ is the $A$-hypergeometric polynomial
\begin{equation}
  \sum_{A u = \beta, u \in {\bf N}_0^n} \frac{1}{u!} x^u, \quad u! = \prod_{i=1}^n u_i!,
  x^u =  \prod_{i=1}^n x_i^{u_i}
\end{equation}
for $\beta \in {\bf N}_0 A=\sum_{i=1}^n {\bf N}_0 a_i$.
It follows from the contiguity relation that we also have
a contiguity relation for derivatives of $Z$
$$ \pd{i} \bullet (s_j \bullet Z)(\beta;x) = (s_j \bullet Z)(\beta-a_i;x) $$
We define the column vector $Y$ by $(s \bullet Z \,|\, s \in S)$.
When $S$ is a basis for both $R'/R'H_A(\beta)$
and $R'/R' H_A(\beta-a_i)$,
it follows from the contiguity relation and the Pfaffian system, 
we have the following identity
\begin{equation} \label{eq:difference_rel}
   Y(\beta-a_i;x) = \pd{i} \bullet Y(\beta;x) = 
    P_i(\beta;x) Y(\beta;x),
\end{equation}
which is called a recurrence relation or a difference Pfaffian system for the vector valued function $Y$ to the direction $a_i$.

We suppose that the value $Y(\cc,x)$ at $\cc=\beta \in {\bf N}_0 A$ and 
$x=X \in {\bf Q}^n $
is given.
Once we construct the difference Pfaffian system,
the value of $Y$ at $\cc=\beta+a_i$ and $x=X$
is obtained by 
\begin{equation}  \label{eq:recc_general}
  P_i(\beta+a_i,x)^{-1} Y(\beta,X)
\end{equation}
when the inverse matrix of $P_i$ exists.
When $\cc=0$, the hypergeometric polynomial $Z(0;x)$ is equal to $1$.
Then, the vector valued function $Y(0;x)$ is $(1, 0, \ldots, 0)^T$.
We apply the recurrence relation (\ref{eq:recc_general}) from $Y(0;x)$ iteratively 
for $i$'s.
Then we can obtain the exact value of $Y(\sum_{m=1}^n h_i a_i)$, $h_i \in {\bf N}_0$
as long as the inverse matrices of $P_i$'s exist
and $S$ is a basis for all $R'/R'H_A(\beta)$.
This method to evaluate hypergeometric polynomials
is called (difference) {\it holonomic gradient method}
((difference) HGM) as an analogy of evaluating
normalizing constants for unnormalized probability distributions
by utilizing holonomic systems of differential equations
\cite{N3OST2}, \cite{HNTT}.

\begin{figure}[tbh]
\setlength\unitlength{1mm}
\begin{picture}(100,40)(0,0)
\put(0,0){\fbox{ideal $I$}}
\put(15,2){\vector(1,0){45}}
\put(23,5){Gr\"obner basis method}
\put(60,0){\fbox{Pfaffian system $P$ of $I$}}
\put(65,10){\vector(0,1){20}}
\put(68,20){Specialization of $\cc$ to $\beta+kH$, $x$ to $X$}
\put(40,33){\fbox{Restriction of $P$ on a line $\cc=\beta+kH$ and $x=X$}}
\put(10,8){\vector(2,1){35}}
\put(5,15){Method of Macaulay type matrices}
\end{picture}
\caption{Restriction to $c=\beta+kH$ and $x=X$}  \label{fig:restriction}
\end{figure}
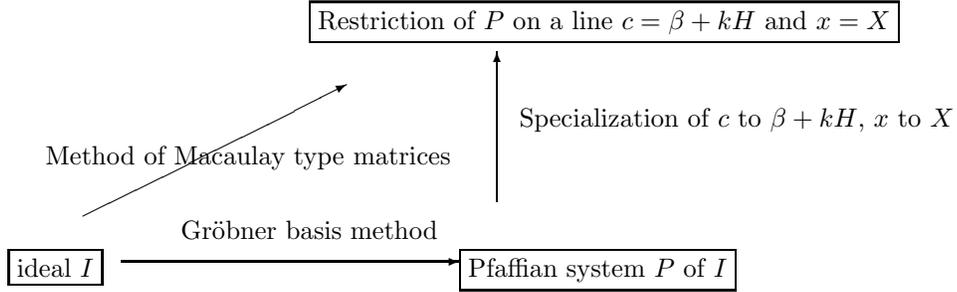

Finding a recurrence relation of $Z$ for indeterminates
$x_1, \ldots, x_n$ and
$\cc_1, \ldots, \cc_d$
is possible by deriving a Pfaffian system by a Gr\"obner basis,
but it 
requires huge computer resouces.
When we want to evaluate $Y(\cc;x)$ at  
a vector of rational numbers $x=X$ 
on a line $\cc= \beta +k H$ parametrized with a parameter $k$
and a direction vector $H$
from the value $Y(\beta;X)$,
we only need the matrix $P_i$ restricted on $x=X$
and $\cc=\beta+kH$.
In order find such restricted $P_i$, we can apply the method of
Macaulay type matrix and avoid Gr\"obner basis computation
for indeterminates $x$ and $\cc$.
This idea is illustrated in the Figure \ref{fig:restriction}

\begin{algorithm}\rm \label{alg:recc-sylvester} (Method of Macaulay type matrices for $A$-hypergeometric systems ) \quad\\  
Input: a matrix $A$, a starting point of the parameter vector 
$\beta \in {\bf N}_0^d$,
a vector of numbers $X \in {\bf Q}^n$,
a basis $S$ of $R_n/R_n H_A(\cc)$ consisting of monomials of $\pd{}$,
a direction $H \in {\bf N}_0^d$ 
such that $H \in \sum_{i=1}^n {\bf N}_0 a_i$. \\
Output: the recurrence relation 
\begin{equation}  \label{eq:recc}
 Y(k-1) = R(k) Y(k) 
\end{equation}
with respect to $k$ for the vector valued hypergeometric functions
$Y(k)=(S\bullet Z)(\beta+kH;X)$. 
\begin{enumerate}
\item Find $h_i \in {\bf N}_0$ such that
$H = \sum_{i=1}^n h_i a_i$ (express the direction $H$ in terms of $a_i$).
This step can be performed by solving the integer program problem
of minimizing $w \cdot h$ under the constraint $Ah=H$, $h \in {\bf N}_0^d$
for a weight vector $w$.
\item Let $G_{I_A}$ be the Gr\"obner basis of the toric ideal $I_A$ in the Algorithm \ref{alg:MacaulayTypeMatrix}.
Let $s'_j$ be the normal form $N(\pd{}^h s_j, G_{I_A})$ of $\pd{}^h s_j$
with respect to the Gr\"obner basis $G_{I_A}$.
Choose sufficiently large number $T$ such that all $s'_j$, $j=1, \ldots, r$
are contained in $M' \cup S$ in the Algorithm \ref{alg:MacaulayTypeMatrix}.
\item Construct the Macaulay type matrix $F'_T$ by the Algorithm
\ref{alg:MacaulayTypeMatrix}.
\item Restrict $F'_T$ to $x=X$ and $\cc=\beta+kH$ and set the obtained
matrix $F''_T$.
If the reduced row echelon form of $F''_T$ contains the element
standing for
\begin{equation}
 s'_i \equiv \sum_{j=1}^r t_{ij} s_j, \quad t_{ij} \in {\bf Q}(k)
\end{equation}
for all $j=1, \ldots, r$,
then go to the step 5
else increase $T$ and go to the step 3.
\item Define the matrix $R(k)$ by $(t_{ij})$.
\end{enumerate}
\end{algorithm}

\begin{example}\rm
Put $A=\left(\begin{array}{cccc}
1&1&1&1 \\
0&1&0&1 \\
0&0&1&1 \\
\end{array}\right)$ and
set the direction $H=(1,1,1)^T$, which stands for $\pd{4}$,
put the starting point
$\beta=(3,2,1)^T$, and put $X = (1,1,1/2,1)$.
As we have seen in Example \ref{ex:matrixE22}, 
we can choose $S=(1,\pd{4})^T$.
Then, we obtain
\begin{eqnarray*}
&&
\left(\begin{array}{c}
Z(\beta-(k-1)H;X)\\
(\pd{4}\bullet Z)(\beta-(k-1)H;X)\\
\end{array}\right) \\
&=&
\left(\begin{array}{cc}
0&1 \\
-2(k+1)(k+2) & 3k+5 \\
\end{array}\right)
\left(\begin{array}{c}
Z(\beta-kH;X)\\
(\pd{4}\bullet Z)(\beta-kH;X)\\
\end{array}\right)
\end{eqnarray*}
by our algorithm.
This recurrence relation holds for any hypergeometric polynomial
for the parameter $\cc=\beta+kH$.
The hypergeometric polynomial $Z(\cc;x)$
is 
$\frac{x_1 x_2 x_4^{k+1}}{(k+1)!} + 
x_2^2 x_3 x_4^k$ 
for 
$\cc=(3+k,2+k,1+k)^T$.
\end{example}

\section{Hilbert Driven Algorithm for $A$-Hypergeometric Ideals}

In the previous section, we discussed a method to evaluate hypergeometric
polynomial by contiguity relations and Macaulay type matrices.
The evaluation by contiguity relations (the difference HGM)
can also be performed if we have a Gr\"obner
basis for the ideal $H_A(\cc)$.
The rank of $R H_A(\cc)$ 
where $\cc_i$'s are  indeterminates is equal to
the normalized volume of $A$
by Adolphson's theorem \cite{adolphson}, then 
the Hilbert driven algorithm by Traverso \cite{traverso}
can be applied to the $A$-hypergeometric
ideal $H_A(c)$.
Let us summarize this method as an algorithm.
\begin{algorithm} \rm \label{alg:hilbgb} (Hilbert driven method for
$A$-hypergeometric systems) \quad \\
Input: a matrix $A$ standing for the $A$-hypergeometric system and 
$\beta \in {\bf N}_0 A$. \\  
Output: recurrence relations (difference Pfaffian systems).
\begin{enumerate}
\item Evaluate the normalized volume of $A$. Put it $r$.
\footnote{The number $r$ can be obtained by evaluating the multiplicity of $I_A$
or by evaluating the volume by geometric methods.}
\item Compute the Gr\"obner basis $G$ of $H_A(\cc)$ in the ring of differential operators
$R={\bf Q}(\cc_1, \ldots, \cc_d,x_1, \ldots, x_n)\langle \pd{1}, \ldots, \pd{n}\rangle$.
We use the Hilbert driven algorithm to avoid unnecessary $S$-pair checks.
In other words, we stop the Buchberger algorithm once the number of standard monomials
of an intermediate Gr\"obner basis equals to $r$.
\item Find a path from $\beta$ to
$\beta'$ which  is near to $0$
by the Algorithm \ref{alg:path}.
\item Let $S$ be the vector of the standard monomials for $G$.
Compute matrices $P_i$ such that $\pd{i} S - P_i S$ modulo $G$ by the normal form algorithm
only for the indexes $i$ which appear in the path.
\item We have recurrences $P_i(\cc+a_i) Y(\cc+a_i) =  Y(\cc)$
on the path.
\end{enumerate}
\end{algorithm}
At each step of applying recurrences, we specialize parameters $c$ and $x$ in $P_i$ to numbers.
It is our heuristic observation that $c_i$ should not  be speicialized to numbers
or parametric polynomials during the Gr\"obner basis computation and 
the normal form computation.
It seems to make the computation slower.

\section{Comparison of Methods to Evalute Hypergeometric Polynomials}

In the latter two sections, we have illustrated two methods
to evaluate $A$-hypergeometric polynomials numerically.
We will compare the two methods and the method by
enumerating the fiber  $\{u \in {\bf N}_0^d \,|\, Au=\beta \}$.

The advantage of the method of Macaulay type matrices
is that the step 4 of the computation
of the reduced row echelon form is performed in the field ${\bf Q}(k)$.
On the other hand, the Gr\"obner basis method needs computation
in the ring ${\bf Q}(\cc,x)\langle \pd{1}, \ldots, \pd{n} \rangle$,
which contains $d+2n$ indeterminates.

All timing data in this section is taken on a machine with Intel Xeon
CPU (2.70GHz) with 256 G memory.
\begin{example} \rm \label{ex:c111c}
Timing data of {\tt test\_c111c\_conti2() } in the package
\cite{ot-hgm-ahg} for Risa/Asir.

Put
$A=
\left(\begin{array}{cccccccc}
 1&  1&  1&  1&  1&  1&  1&  1 \\
0&  1& 0& 0&  1&  1& 0&  1 \\
0& 0&  1& 0&  1& 0&  1&  1 \\
0& 0& 0&  1& 0&  1&  1&  1 \\
\end{array}\right)
$,
$\beta=(3,2,1,1)^T$. \\
$X= (1,1/2,1/3,2/3,1,1,1,1)$
$S=(1,\pd5,\pd6,\pd7,\pd8,\pd8^2)^T$.

Our benchmark problem is to 
get the value of 
$S \bullet Z $ at $x=X$
and $c=\beta+k(3,1,1,1)^T$.
The contiguity for $(1,1,1,1)^T$ is applied to obtain the value
at $\beta+k(1,1,1,1)^T$ and next
the contiguity for $(1,0,0,0)^T$ is applied to obtain the value
at $\beta+k(3,1,1,1)^T$ from the value at
$\beta+k(1,1,1,1)^T$.

THe following table is timing data of the method of Macaulay type
matrices. \\
\begin{tabular}{c|c}
k  &  Time (second) \\ \hline
0&  1.45 \\
 10&  1.48 \\
 20&  1.70 \\
 30&  1.79 \\
 40&  1.89 \\
 50&  1.99 \\
 60&  2.16 \\
 70&  2.35 \\
 80&  2.58 \\
 90&  2.82 \\
 100&  3.17 \\
\end{tabular}

\noindent
The timing for the case $k=0$ is the time to construct recurrence relations
for the direction $(1,1,1,1)^T$ and $(1,0,0,0)^T$.
When $k=10$, 
our implementation outputs
\begin{center}
{\tt Val=[30318066527332447242457/89619251224349337722522492794306560000, ...]}.
\end{center}
From the output,
we obtain, e.g., values 
\begin{eqnarray*}
Z&=&30318066527332447242457/89619251224349337722522492794306560000 \\
 &=&3.38\ldots \times 10^{-16} \\
E[U_8] &= &\pd{8}\bullet Z/Z = 52047189429143224956864/30318066527332447242457 \\
 &=& 1.71\ldots 
\end{eqnarray*}
where $E[U_8]$ is the expectation of the random variable $U_8 \in {\bf N}_0^8$
which satisfies $AU=\beta+k(3,1,1,1)^T$.
We compare these data with other two methods for exact evaluation.
First one is an exhaustive enumeration of $U \in {\bf N}_0^8$
satisfying $A U = \beta+k(3,1,1,1)^T$.
It can be easily enumerated for this $A$ by a nested loop of four {\tt for} statements,
because when non-negative integers $U_5, U_6, U_7, U_8$ are given, 
other $U_i$'s are determined uniquely as integers and the admissible range of
$U_5, \ldots, U_8$ can be described in terms of $k$.
Moreover coefficients can easily be evaluated by recurrences.

\noindent
\begin{tabular}{c|c|c}
k  &  Time (second) &number of fibers\\ \hline
0& 0&  5 \\
 10&  0.0200&  1946 \\
 20&  0.132&  18436 \\
 30&  0.476&  76976 \\
 40&  1.42&  220066 \\
 50&  3.70&  505206 \\
 60&  7.68&  1004896 \\
 70&  15.9&  1806636 \\
 80&  30.9&  3012926 \\
 90&  56.0&  4741266 \\
 100&  88.7 &  7124156 \\
\end{tabular}

\begin{figure}[tb]
\begin{center}
\includegraphics[width=7cm]{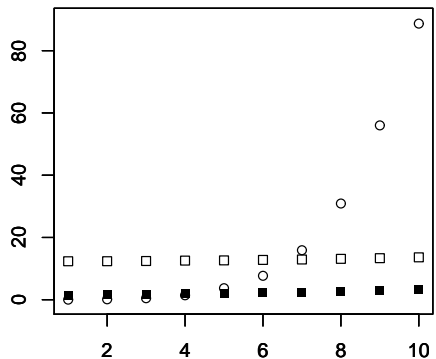}
\end{center}
\caption{black square=Macaulay type matrix, 
   white square=Hilbert driven GB, white circle=series, horizontal=$k/10$, vertical=seconds}  \label{fig:timing3}
\end{figure}

The last table is timing data of  evaluating the normalizing constant
in the benchmark problem
by recurrence relations
derived from Pfaffian operators obtained by
the algorithm \ref{alg:hilbgb}.
The Hilbert driven Gr\"obner basis computation in the step 2 takes 
5.50 seconds on the yang package for the Risa/Asir.
If we do not use the Hilbert driven method in the step 2,
it takes 2069.78 seconds on the yang package.
In our benchmark problem, we may only obtain $P_1$ and $P_8$.
It takes 6.80 seconds to obtain them in the step 3.  \\
\begin{tabular}{c|c|c}
$k$ & Time Recc (second) & Time Recc $+$ GB (second) \\ \hline
0&  0.02&  12.31 \\
 10&  0.06&  12.35 \\
 20&  0.072&  12.37 \\
 30&  0.14&  12.43 \\
 40&  0.27&  12.56 \\
 50&  0.34&  12.64 \\
 60&  0.49&  12.78 \\
 70&  0.62&  12.92 \\
 80&  0.83&  13.13 \\
 90&  1.05&  13.34 \\
 100&  1.32&  13.62 \\
\end{tabular} \\
In the table, ``Time Recc'' is the time to apply the recurrence relations
to obtain the value of the normalizing constant.
``Time Recc $+$ GB'' is the sum of the ``Time Recc'' and the time of
the algorithm \ref{alg:hilbgb}.
\end{example}

The Figure \ref{fig:timing3} tells us a strategy to evaluate
$A$-hypergeometric polynomials numerically.
When $\beta$ is small, then the enumeration of the fibers will be
the best method.
When $\beta$ is large, then the method of Macaulay type matrix
will be the best choice.
When we need to evaluate the hypergeometric polynomial
for several $\beta$'s, 
then the method of Gr\"obner basis
by the Hilbert driven algorithm will be the best choice.

\section{Finding a Path to Apply for Recurrence Relations}

Suppose that $\beta \in {\bf N}_0 A$.
We want to find a path in ${\bf N}_0 A$ from $\beta$ to 
$\beta'$ closer to $0$

\comment{Algorithm of {\tt find\_path} for $A$ and $\beta$
is in Prog/note-t.tex. Correctness is not proved.}
\begin{algorithm} \rm \quad \\
Input: $A$ and $\beta \in {\bf N}_0A$. \\
Output: a path to $\beta' \in {\bf N}_0^d$  \label{alg:path}
\begin{enumerate}
\item Express $\beta$ as 
$ \beta = n_1 a_{i_1} + \cdots + n_k a_{i_k}$,  $n_i \in {\bf Z}_{>0}$
by reducing $\pd{}^\beta$ by a Gr\"obner basis of $I_A$.
\comment{{\tt tk\_hgpoly.optip}}.
\item Find the maximum in the set
\begin{eqnarray*}
 \{ m &|& \mbox{path $\beta-\gamma_u \rightarrow \beta-\gamma_u-m a_{i_j}$} \\
 &&
\mbox{ lies in ${\bf N}_0 A$ }
 \mbox{ for all $\gamma_u = \sum u_k a_k$, $\pd{}^u \in S$ } 
\mbox{ and }
 m \leq n_j \}
\end{eqnarray*}
When the set is empty or $\beta=0$, return the path and 
$\beta$ as the terminating point $\beta'$.
\item Add the pair $(i_j,m)$ to the path. 
Put $n_j = n_j - m$ and $\beta = \beta-m a_{i_j}$.
Go to the step 2.
\end{enumerate}
\end{algorithm}
\noindent
The definition that 
``path $\beta-\gamma_u \rightarrow \beta-\gamma_u-m a_{i_j}$
lies in ${\bf N}_0 A$''
is that
$\beta-\gamma_u - a_{i_j} \in {\bf N}_0A$,
$\beta-\gamma_u - 2a_{i_j} \in {\bf N}_0A$,
$\ldots$,
$\beta-\gamma_u - ma_{i_j} \in {\bf N}_0A$.


Let $S$ be a basis of $R/RH_A(c)$
consisting of monomials in $\pd{i}$'s
and $\pd{i} S - P_i S$, $i=1, \ldots, n$ is the Pfaffian operator for $S$.
The singularity polynomial of the Pfaffian operator is the least common
multiple of the denominator polynomials of the elements of the $P_i$'s.
It is a polynomial in ${\bf C}[c,x]$.
The basis $S$ is called a {\it good basis} for ${\bf N}_0A$
when the following two conditions are satisfied.
\begin{enumerate}
\item The singularity polynomial for the Pfaffian operator does not contain
the variables $c_i$'s.
\item $S$ is a basis of $R'/R'H_A(\beta)$,
$R'={\bf C}(x)\langle \pd{1}, \ldots, \pd{n} \rangle$
for all $\beta \in {\bf N}_0 A$.
\end{enumerate} 
Note that when $A$ admits a rank jumping parameter $\beta \in {\bf N}_0 A$
\cite{MMW},
there exists no good basis.

The following theorem shows that the algorithm \ref{alg:path} 
and HGM reduce the evaluation
of $Z(\beta;x)$ to $Z(\beta';x)$
where $\beta'$ closer to $0$
when the condition of the theorem holds.

\begin{theorem}
When $A$  is normal and $S$ is a good basis,
then the matrix $P_i(\beta+a_i;x)$  has the inverse for $\beta$
satisfying  $\beta-\sum u_i a_i \in {\bf N}_0 A$
for all $\pd{}^u \in S$ 
and for any $x$ out of a measure zero set.
\end{theorem}

{\it Proof}\/.
We will construct the inverse matrix of $P_i$.
We denote by $M_A(\beta)$ the left $D_n$-module $D_n/D_n H_A(\beta)$.
Since $A$ is normal, it follows from Saito's isomorphism
\cite{Siso} that $M_A(\beta)$ and $M_A(\beta+a_i)$ are isomorphic
as the left $D_n$-module
when $\beta \in {\bf N}_0 A$.
The isomorphism is given by
$$ M_A(\beta) \ni \ell \mapsto \ell \pd{i} \in M_A(\beta+a_i). $$
In general, they are isomorphic when
$\beta+a_i \not\in V(B_i)$, which is the zero set of the b-ideal 
for the direction $i$ \cite{SST2}.
When $A$ is normal, we have $\beta +a _i \not\in V(B_i)$
for all $i$ and $\beta \in {\bf N}_0 A$.
\comment{ Consider $\ell_i \pd{i} -1 \in H_A(c+a_i) \cap K[c]$}
\noindent
Note that we have the induced isomorphism
$${\rm Hom}_{D_n}(M_A(\beta+a_i), {\cal O}) \ni f 
\mapsto 
\pd{i} \bullet f \in
{\rm Hom}_{D_n}(M_A(\beta), {\cal O}) .
$$
Let $\pd{}^u$ be an element of $S$.
When $\beta-\sum u_k a_i \in {\bf N}_0 A$,
there exists an operator $\ell_{iu} \in D_n$ such that
the following diagram of left $D_n$-modules commutes
and all arrows are isomorphisms.
\begin{center}
\xymatrix{
& M_A(\beta-\sum u_k a_k) \ar@{->}^{\pd{}^u}[d] \ar@{<-}^{\ell_{iu}}[r] 
& M_A(\beta+a_i-\sum u_k a_k) \ar@{->}^{\pd{}^u}[ddl] \\
& M_A(\beta) \ar@{->}^{\pd{i}}[d] & \\
& M_A(\beta+a_i) & 
}
\end{center}
\noindent
Since the diagram commutes, we have
$$ \ell_{iu} \pd{}^u \pd{i} - \pd{}^u = 0 \ \mbox{ in }  M_A(\beta+a_i), $$
which implies
$ \ell_{iu} \pd{}^u \pd{i} - \pd{}^u  \in D_n H_A(\beta+a_i)$.
Let $f$ be a local solution of $M_A(\beta)$,
which can be regarded as an element of
${\rm Hom}_{D_n}(M_A(\beta),{\cal O})$.
Consider the column vector $Y_f(\beta) = ( s \bullet f \,|\, s \in S)$.
Take a local solution $g$ in ${\rm Hom}_{D_n}(M_A(\beta+a_i),{\cal O})$.
We have
$\ell_{iu} \pd{}^u \pd{i} \bullet g - \pd{}^u \bullet g = 0$.
The function $f = \pd{i} \bullet g$ is a solution of $M_A(\beta)$.
Then, we have 
$Y_g(\beta+a_i) =
(\pd{}^u \bullet g \,|\, \pd{}^u \in S) = (\ell_{iu} \pd{}^u \bullet f \,|\, \pd{}^u \in S)
$.
Suppose that $\ell_{iu} \pd{}^u = \sum c_v(x) \pd{}^v$.
Since $S$ is a good basis, $\pd{}^v$ can be expressed as a linear 
combination of $S$ with coefficients in ${\bf C}(x)$
modulo $R' H_A(\beta)$.
Therefore, there exists a matrix $Q_i$ with rational function entries
satisfying
$Y_g(\beta+a_i) = Q_i(x) Y_f(\beta)$.
We note that $Q_i$ does not depend on the choice of the function $g$.
It follows from the contiguity relation that we have
$ Y_f(\beta) = P_i(\beta+a_i,x) Y_g(\beta+a_i)$.
Hence, we have
$Y_g(\beta+a_i) = Q_i(x) P_i(\beta+a_i,x) Y_g(\beta+a_i)$.
Make $g$ run over a solution basis.
Then, we can see that  
$Q_j(x) P_i(\beta+a_i,x) = E$ out of an analytic set in the $x$-space.
The analytic set defined by a non-zero holomorphic function is a
measure zero set.
A countable union of measure zero sets is a measure zero set.
\qed

Note that we do not have an example of  $S$
which is not good when $A$ is normal.
It is an interesting question to find such example
or to prove the existance of a good basis.

\section{Reduction to Non-Negative $a_i$'s}

We have supposed $a_i \in {\bf N}_0^d$ in the previous sections.
We assume that $a_i \in {\bf Z}^d$
and there exists a linear form $h \in {\bf R}^n$ such that
$h(a_i) = 1$ for all $i$'s
in this section.
We will explain a method to reduce this case to  the case
$a_i \in {\bf N}_0^d$.

We define the vector $\pp \in {\bf N}_0^n$ by
$$ \pp_i  = - {\rm min}_j (a_{ij},0).
$$
When $h(\pp) = -1$, we choose a unit vector $e$ which is not
orthogonal to $h$ and rewrite $\pp$ by $\pp+e$.
Then, we can assume that $h(\pp) \not= -1$.
Define the new matrix $A'$ by
$(a_1+\pp, \ldots, a_n+\pp)$.
The toric ideals for $A$ and for $A'$ agree,
e.g., \cite[Lemma 1.5.10]{dojo}.

Let us discuss on a relation between solutions of
the $H_A$ and $H_{A'}$.
When $f$ is a solution of $H_A(\beta)$, 
we have
\[
\sum_{j=1}^n a_{ij} x_j \pd{j} f = \beta_i f .
\]
Multiplying $h_i$ and adding with respect to $i$, we obtain
\[
\sum_{j=1}^n x_j \pd{j} f = h(\beta) f
\]
by utilizing $h(a_j)=1$.
Adding 
$ p_k (\sum_{j=1}^n x_j \pd{j} f )= h(\beta) p_k f $
to the $k$-th equation $E_kf = \beta_k f$,
we obtain
$$ \sum (a_{kj} +p_k) x_j  \pd{j} f = (\beta_k+ h(\beta) p_k ) f $$
Hence, the solutions of $H_A(\beta)$ are solutions of
$H_{A'}(\beta')$
where 
$\beta'=\beta+h(\beta)c$.
In particular, the hypergeometric polynomial for $H_{A}(\beta)$ 
is the hypergeometric polynomial for $H_{A'}(\beta')$.

\end{document}